\documentclass[journal,onecolumn]{IEEEtran}
\usepackage{amsmath,amsfonts,amssymb}
\usepackage{algorithmic}
\usepackage{algorithm}
\usepackage{array}
\usepackage[caption=false,font=normalsize,labelfont=sf,textfont=sf]{subfig}
\usepackage{textcomp}
\usepackage{stfloats}
\usepackage{url}
\usepackage{verbatim}
\usepackage{graphicx}
\usepackage{cite}
\usepackage{xcolor}
\newtheorem{example}{Example}
\newtheorem{theorem}{Theorem}
\newtheorem{lemma}{Lemma}
\newtheorem{construction}{Construction}
\newtheorem{remark}{Remark}

\begin{document}

\title{$t$-Deletion-$s$-Insertion-Burst Correcting Codes}

\author{
\textbf{Ziyang Lu} and \textbf{Yiwei Zhang}
\thanks{The authors are with Key Laboratory of Cryptologic Technology and Information Security, Ministry of Education,
and also with School of Cyber Science and Technology, Shandong University, Qingdao, Shandong, 266237, China.
Emails: zylu@mail.sdu.edu.cn, ywzhang@sdu.edu.cn.}
\thanks{Part of this work was presented at IEEE International Symposium on Information Theory (ISIT), 2022.}
\thanks{Research supported by National Key Research and Development Program of China under Grant Nos. 2020YFA0712100 and 2021YFA1001000,
National Natural Science Foundation of China under Grant No. 12001323, and Shandong Provincial Natural Science Foundation under Grant No. ZR2021YQ46.}
}

\maketitle

\begin{abstract}
Motivated by applications in DNA-based storage and communication systems,
we study deletion and insertion errors simultaneously in a burst.
In particular, we study a type of error named $t$-deletion-$s$-insertion-burst ($(t,s)$-burst for short)
which is a generalization of the $(2,1)$-burst error proposed by Schoeny {\it et. al}.
Such an error deletes $t$ consecutive symbols and inserts an arbitrary sequence of length $s$ at the same coordinate.
We provide a sphere-packing upper bound on the size of binary codes that can correct a $(t,s)$-burst error,
showing that the redundancy of such codes is at least $\log n+t-1$.
For $t\geq 2s$, an explicit construction of binary $(t,s)$-burst correcting codes with redundancy $\log n+(t-s-1)\log\log n+O(1)$ is given.
In particular, we construct a binary $(3,1)$-burst correcting code with redundancy at most $\log n+9$, which is optimal up to a constant.
\end{abstract}


\section{Introduction}
DNA-based storage is a promising direction for future data storage due to its advantages such as storage density and durability\cite{Dong-DNABcakground-NSR20}.
Due to the error behavior in DNA sequences, codes correcting deletions and insertions have recently attracted significant attention.
Meanwhile, synchronization loss occurs due to timing uncertainty in communication and storage systems\cite{Sala-Synchronizing-TC16,Dolecek-Synchronizing-IT07},
which also leads to deletions and insertions.

The study of error-correcting codes against deletions and insertions dates back to 1965,
when Levenshtein\cite{Levenshtein-Deletion-1965} showed that the Varshamov-Tenengol'ts (VT) code\cite{VTcode-AT65} can correct a single deletion.
In the same work, Levenshtein proved the equivalence between $t$-deletion-correcting codes and $t$-insertion-correcting codes,
and showed that the redundancy of a $t$-deletion-correcting code is asymptotically at least $t\log n$.
It was not until recent years that deletion correcting codes were reconsidered due to advances in DNA storage.
The general problem of correcting $t$ arbitrary deletions has been considered in a series of works\cite{Brakensiek-kDeletion-IT18,Tian-PolorDeletion-ISIT18,Hanna-GCcode-IT19,Gabrys-2Del-IT19,Guruswami-2Del-IT21,Sima-kDeletion-IT21,Sima-tDeletion-ISIT20}.
The state-of-the-art result of them is the one from\cite{Sima-tDeletion-ISIT20},
which constructed a $t$-deletion-correcting code with redundancy $4t\log n+o(\log n)$.

Both in DNA-based storage and communication systems,
the deletion and insertion errors tend to occur in bursts (i.e., consecutive errors) \cite{Yazdi-DNABackground-15}.
Therefore, it is of great significance to design codes capable of correcting a burst of deletions/insertions.
A code that can correct {\it exactly} $t$ consecutive deletions is called a \textit{$t$-burst-deletion-correcting code},
and a code that can correct {\it at most} $t$ consecutive deletions is called a \textit{$t^{\le}$-burst-deletion-correcting code}.
In 1970, Levenshtein\cite{Levenshtein-2Burst-1970} constructed a $2^{\le}$-burst-deletion-correcting code,
and provided asymptotic bounds on $t$-burst-correcting codes indicating that the least redundancy to correct $t$ consecutive deletions is asymptotically $\log n+t-1$.
Cheng \textit{et al.}\cite{cheng-Burst-ISIT14} provided three constructions of $t$-burst-deletion-correcting codes,
among which the least redundancy is $t\log(\frac{n}{t}+1)$.
Schoeny \textit{et al.}\cite{Schoeny-Burst-IT17} constructed a $t$-burst-deletion-correcting code by combining the VT code with constrained coding and utilizing the shifted VT (SVT) code, which has redundancy $\log n+(t-1)\log\log n+t-\log t$.
As for $t^{\le}$-burst-deletion-correcting codes, the construction of Schoeny \textit{et al.}\cite{Schoeny-Burst-IT17} has redundancy $(t-1)\log n+(\binom{t}{2}-1)\log\log n+\binom{t}{2}+\log\log t$ and then Gabrys \textit{et al.} \cite{Gabrys-Damerau-IT18} reduced the redundancy to $\lceil\log t\rceil\log n+(\frac{t(t+1)}{2}-1)\log\log n+O(1)$. The current best result for $t^{\le}$-burst-deletion-correcting codes is from Lenz and Polyanskii\cite{Lenz-BurstVariLen-ISIT20} with redundancy $\log n+\frac{t(t+1)}{2}\log\log n+O(1)$. In addition, the model of a burst of $t$ non-consecutive deletions (some $s\le t$ deletions occur within a block of size $t$) is also considered in \cite{Schoeny-Burst-IT17,Bitar-BurstNConsec-arxiv21}.

In addition to deletion or insertion error, substitution error is the most widely considered type of error in classic coding theory.
In current DNA storage technology, ultimately one needs to design error-correcting codes against a combination of deletion, insertion, and substitution errors,
which is by now a difficult problem\cite{Smagloy-delsub-ISIT20,Cai-Edit-21,Gavrys-DelSubTransp-arxiv21,Song-MultiDelSub-ISIT21,Song-ListDelSub-Arxiv22}.

In \cite{Schoeny-Burst-IT17} Schoeny \textit{et al.} proposed a type of error called a $(2,1)$-burst,
which deletes two consecutive symbols and then inserts one symbol at the same coordinate.
Codes against a $(2,1)$-burst error were applied in their model of burst non-consecutive deletions \cite{Schoeny-Burst-IT17}.
In this paper, we generalize the error type to a \textit{$t$-deletion-$s$-insertion-burst} ({\it $(t,s)$-burst} for short),
which deletes a burst of $t$ consecutive symbols and then inserts an arbitrary sequence of length $s$ at the same coordinate.
Despite being naturally a mixture of deletions and insertions, a $(t,s)$-burst can also be seen as a mixture of substitutions and deletions/insertions.
For example, when $t\geq s$, a $(t,s)$-burst can be also seen as deleting a burst of $t-s$ consecutive symbols and then substituting a subset of the next $s$ consecutive symbols. A code that can correct a $(t,s)$-burst error will be called a \textit{$t$-deletion-$s$-insertion-burst correcting code} ($(t,s)$-burst correcting code for short), and is the main objective of this paper.

The rest of the paper is organized as follows.
In Section \ref{sec:pre}, we give the definitions and some notations used throughout the paper and review some previous results that will be used in our constructions.
In Section \ref{sec:equiv}, we prove the equivalence between $(t,s)$-burst correcting codes and $(s,t)$-burst correcting codes.
A sphere-packing upper bound on the size of $(t,s)$-burst correcting codes is given in Section \ref{sec:upper}, which leads to a lower bound of the code redundancy.
In Section \ref{sec:t-s-constr}, we present a $(t,s)$-burst correcting code with redundancy $\log n+(t-s-1)\log\log n+O(1)$ for $t\geq2s$.
In particular, we present an almost optimal $(3,1)$-burst correcting code in Section \ref{sec:3-1-constr}.
Lastly, Section \ref{sec:concl} concludes the paper and discusses some future directions.

\section{Preliminaries and Related Work}\label{sec:pre}
\subsection{Notations and Definitions}
Let $\mathbb{F}_2^n$ be the set of binary sequences of length $n$.
In this paper, a sequence $\boldsymbol x\in\mathbb{F}_2^n$ is denoted as either $x_1x_2x_3\ldots x_n$ or $(x_1,x_2,x_3,\ldots,x_n)$.
For integers $i\le j$, the {\it length} of an interval $[i,j]\triangleq\{i,i+1,\dots,j\}$ is defined as $j-i+1$.

A {\it run} of $\boldsymbol x=(x_1,x_2,\dots,x_n)$ is a maximal consecutive subsequence of the same symbol.
For binary sequences, a run is a maximal consecutive 0s or 1s.
Let $R(\boldsymbol x)$ denote the \textit{run sequence} of $\boldsymbol x$, and the $i$th coordinate of $R(\boldsymbol x)$ is denoted as $R(\boldsymbol x)_i$,
known as the \textit{run index} of $x_i$.
Here $R(\boldsymbol x)_i=t$ indicates that $x_i$ lies in the $t$th run of $\boldsymbol x$, where $t$ is counted starting from zero.
Let $Rsyn(\boldsymbol x)=\sum_{1\le i \le n} R(\boldsymbol x)_i$ and let $r(\boldsymbol x)$ be the total number of runs in $\boldsymbol x$.
For example, if $\boldsymbol x=1101110000$, then `11', `0', `111', `0000' are the four runs,
$R(\boldsymbol x)=0012223333$, $Rsyn(\boldsymbol x)=19$, and $r(\boldsymbol x)=4$.

A \textit{$t$-deletion-$s$-insertion-burst} ({\it $(t,s)$-burst} for short) error over $\boldsymbol x$ is a type of error which deletes $t$ consecutive symbols from $\boldsymbol x$ and inserts an arbitrary sequence of length $s$ at the same coordinate.
That is, a $(t,s)$-burst error over $\boldsymbol x=(x_1,x_2,\dots,x_n)$ starting at the $i$th coordinate, $1\leq i \leq n-t+1$, will result in $(x_1,\dots,x_{i-1},y_1,\ldots,y_s,x_{i+t},\dots,x_n)$, where $(x_i,\dots,x_{i+t-1})$ is deleted and $(y_1,\ldots,y_s)$ is the inserted sequence.
Continuing with the example above for $\boldsymbol x=1101110000$,
if a $(3,1)$-burst error starts at the very beginning then one gets either $01110000$ or $11110000$,
and if a $(1,3)$-burst error starts at the very beginning then one gets $y_1y_2y_3101110000$ where $y_1y_2y_3$ is the inserted sequence.

The \textit{$(t,s)$-burst ball} of $\boldsymbol x$ is the set of all possible sequences obtained by a $(t,s)$-burst error over $\boldsymbol x$,
denoted as $B_{t,s}(\boldsymbol x)$.

\begin{example}
Suppose $\boldsymbol x=101000111$,  $t=4,s=1$, then
\begin{equation*}
   B_{4,1}(\boldsymbol x)=  \{000111,100111,110111,101111,
      101011,101001,101000\}.
\end{equation*}
\end{example}

A code $\mathcal{C}\subseteq\mathbb{F}_2^n$ is called a \textit{$(t,s)$-burst correcting code} if it can correct a $t$-deletion-$s$-insertion-burst error.
That is, for any two distinct codewords $\boldsymbol c_1,\boldsymbol c_2\in\mathcal{C}$, $B_{t,s}(\boldsymbol c_1)\cap B_{t,s}(\boldsymbol c_2)=\varnothing$.
The redundancy of such a code is given by $n-\log\vert\mathcal{C}\vert$.

\subsection{Related Work}
In this subsection we briefly review some useful results about codes correcting deletions or insertions.

Define the {\it VT syndrome} as $VT(\boldsymbol x)=\sum_{i=1}^{n}ix_i$.
For $a\in\mathbb{Z}_{n+1}$, the Varshamov-Tenengol'ts (VT) code\cite{VTcode-AT65}
$$VT_a(n)=\{\boldsymbol x\in\mathbb{F}_2^n: VT(\boldsymbol x)\equiv a\pmod{n+1}\}$$
can correct a single deletion. Levenshtein \cite{Levenshtein-Deletion-1965} proved that the optimal redundancy of $t$-deletion-correcting codes is between $t\log n$ and $2t\log n$, so the VT code is optimal for a single deletion.

In 1970, Levenshtein\cite{Levenshtein-2Burst-1970} provided a code that can correct a burst of at most 2 deletions as follows:
\begin{equation}\label{Equ:Lev2burst}
L_a(n)=\{\boldsymbol x\in\mathbb{F}_2^n: Rsyn(0\boldsymbol x)\equiv a\pmod{2n}\},
\end{equation}
where $a\in\mathbb{Z}_{2n}$ and the function $Rsyn(\cdot)$ is applied on $0\boldsymbol x$,
the concatenation of a single 0 and the sequence $\boldsymbol x$.

For the $t$-burst-deletion-correcting codes, Cheng \textit{et al.} \cite{cheng-Burst-ISIT14} proposed a framework which represents $\boldsymbol x$ as a $t\times \frac{n}{t}$ array (throughout the paper, in an array representation we always assume that the number of rows is a divisor of $n$) and applies some constraints for each row:
\begin{equation*}
A_t(\boldsymbol x)=
  \begin{bmatrix}
    x_1 & x_{t+1} & \cdots & x_{n-t+1} \\
    x_2 & x_{t+2} & \cdots & x_{n-t+2} \\
    \vdots & \vdots & \ddots & \vdots \\
    x_t & x_{2t} & \cdots & x_n
  \end{bmatrix}.
\end{equation*}
In their constructions the least redundancy is $t(\log(\frac{n}{t}+1))$.

Schoeny \textit{et al.}\cite{Schoeny-Burst-IT17} followed the framework as in \cite{cheng-Burst-ISIT14} to represent a sequence as the array above.
In their construction the first row is encoded by a variation of VT code with additional run-length constraints,
and the other rows apply a code called the shifted VT code (SVT):
\begin{equation*}
     SVT(n;a,b,P)=\Big\{  \boldsymbol x\in\mathbb{F}_2^n: VT(\boldsymbol x)\equiv a\pmod{P},
        \sum_{i=1}^{n}x_i\equiv b\pmod{2}\Big\}
\end{equation*}
which can correct a single deletion given the additional knowledge of the erroneous coordinate within an interval of length $P$ (this knowledge is derived from decoding the first row). By choosing $P$ as $\log(n/t)+2$, the redundancy of their codes is at most $\log n+(t-1)\log \log n+t-\log t$.

In the same paper\cite{Schoeny-Burst-IT17}, Schoeny \textit{et al.} constructed a $(2,1)$-burst correcting code,
which was then applied to correcting a burst of $t$ non-consecutive deletions:
\begin{equation}\label{Equ:2-1-burst}
     \mathcal{C}_{2,1}(n;a,b)=\Big\{  \boldsymbol x\in\mathbb{F}_2^n: VT(\boldsymbol x)\equiv a\pmod{2n-1},
        \sum_{i=1}^{n}x_i\equiv b\pmod{4}\Big\}.
\end{equation}

\subsection{Summary of Our Contributions}
In this paper, we first prove the equivalence of $(t,s)$-burst correcting codes and $(s,t)$-burst correcting codes.
Next we compute the size of the error ball for the $(t,s)$-burst error.
Unlike most situations in analyzing deletions/insertions, for $(t,s)$-burst the size of the error ball is a constant independent of the center and thus it leads to a neat sphere-packing type upper bound, implying that the redundancy of $(t,s)$-burst correcting codes is at least $\log n+t-1$.
As for constructions, we can focus only on the case $t\geq s$ due to the aforementioned equivalence.
For $t\geq2s$, we give a general construction with redundancy $\log n+(t-s-1)\log \log n+O(1)$.
In particular, we construct a $(3,1)$-burst correcting code with redundancy at most $\log n+9$, which is optimal up to a constant.

\section{Equivalence of $(t,s)$-Burst and $(s,t)$-Burst}\label{sec:equiv}
In this section, we prove the equivalence of $(t,s)$-burst correcting codes and $(s,t)$-burst correcting codes.
Levenshtein \cite{Levenshtein-2Burst-1970} proved the equivalence of $t$-deletion-correcting codes and $t$-insertion-correcting codes.
Schoeny \textit{et al.}\cite{Schoeny-Burst-IT17} proved the equivalence of codes against a burst of $t$ deletions and a burst of $t$ insertions.
Using a similar idea, we have the following theorem.
\begin{theorem}
Let $t\geq s$. A code $\mathcal{C}$ is a $(t,s)$-burst correcting code \textit{if and only if} it is an $(s,t)$-burst correcting code.
\end{theorem}
\begin{IEEEproof}
We only prove the `only if' part. The `if' part can be proved analogously.
Suppose $\mathcal{C}$ is a $(t,s)$-burst correcting code but not an $(s,t)$-burst correcting code.
Then, there exist two distinct codewords $\boldsymbol x,\boldsymbol y\in\mathcal{C}$,
such that $B_{s,t}(\boldsymbol x)\cap B_{s,t}(\boldsymbol y)$ is nonempty and thus contains some $\boldsymbol z\in\mathbb{F}_2^{n+t-s}$.
Assume that $\boldsymbol z$ is obtained by deleting $(x_i,\ldots,x_{i+s-1})$ and inserting $(a_1,\ldots,a_t)$ at the $i$th coordinate of $\boldsymbol x$,
and is also obtained by deleting $(y_j,\ldots,y_{j+s-1})$ and inserting $(b_1,\ldots,b_t)$ at the $j$th coordinate of $\boldsymbol y$.
Without loss of generality, we assume $i\leq j$. Then, $\boldsymbol z$ will have the following two representations:
  \begin{equation*}
  \begin{split}
   & \boldsymbol z^1=(x_1,\ldots,x_{i-1},a_1,\ldots,a_t,x_{i+s},\ldots,x_n),\\
   & \boldsymbol z^2=(y_1,\ldots,y_{j-1},b_1,\ldots,b_t,y_{j+s},\ldots,y_n).
  \end{split}
  \end{equation*}
  \begin{enumerate}
    \item If $j\geq i+t$, i.e., the coordinates of deletions of $\boldsymbol y$ are disjoint with the coordinates of insertions of $\boldsymbol x$.
    By comparing $\boldsymbol z^1$ and $\boldsymbol z^2$ we have
        \begin{align*}
         (x_1,\ldots,x_{i-1})=(y_1,\ldots,y_{i-1})&,  (a_1,\dots,a_t)=(y_i,\dots,y_{i+t-1}), \\
         (x_{i+s},\ldots,x_{j+s-t-1})=(y_{i+t},\ldots,y_{j-1}), (x_{j+s-t},\dots,x_{j+s-1})&=(b_1,\dots,b_t), (x_{j+s},\ldots,x_{n})=(y_{j+s},\ldots,y_{n}).
        \end{align*}
    Therefore, if we delete $(x_{j+s-t},\ldots,x_{j+s-1})$ from $\boldsymbol x$ and insert $(y_j,\ldots,y_{j+s-1})$ at this coordinate,
    then we get $$(x_1,\ldots,x_{j+s-t-1},y_j,\ldots,y_{j+s-1},x_{j+s},\ldots,x_n).$$
    If we delete $(y_i,\ldots,y_{i+t-1})$ from $\boldsymbol y$ and insert $(x_i,\ldots,x_{i+s-1})$ at this coordinate, then we get $$(y_1,\ldots,y_{i-1},x_i,\ldots,x_{i+s-1},y_{i+t},\ldots,y_n).$$
    From the equations above, we have
        \begin{equation*}
           (x_1,\ldots,x_{j+s-t-1},y_j,\ldots,y_{j+s-1},x_{j+s},\ldots,x_n) = (y_1,\ldots,y_{i-1},x_i,\ldots,x_{i+s-1},y_{i+t},\ldots,y_n).
        \end{equation*}
    Thus, $B_{t,s}(\boldsymbol x)\cap B_{t,s}(\boldsymbol y)\neq\varnothing$,
    which is a contradiction to the hypothesis that $\mathcal{C}$ is a $(t,s)$-burst correcting code.

    \item If $j\leq i+t-1$, then the coordinates of deletions of $\boldsymbol y$ and the coordinates of insertions of $\boldsymbol x$ have some intersection.
    We still have
        \begin{equation*}
         (x_1,\ldots,x_{i-1})=(y_1,\ldots,y_{i-1}),
         (x_{j+s},\ldots,x_{n})=(y_{j+s},\ldots,y_{n}).
        \end{equation*}
    In this case, if we delete $(x_{j-t+1},\ldots,x_{j})$ from $\boldsymbol x$ and insert $(y_{j-t+1},\ldots,y_{j-t+s})$ at this coordinate,
    then we get $$(x_1,\ldots,x_{j-t},y_{j-t+1},\ldots,y_{j-t+s},x_{j+1},\ldots,x_n).$$
    For $\boldsymbol y$, if we delete $(y_{j-t+s+1},\ldots,y_{j+s})$ and insert $(x_{j+1},\ldots,x_{j+s})$ at this coordinate,
    then we get $$(y_1,\ldots,y_{j-t+s},x_{j+1},\ldots,x_{j+s},y_{j+s+1},\ldots,y_n).$$
    Since $j\leq i+t-1$, we have $(x_1,\ldots,x_{j-t})=(y_1,\ldots,y_{j-t})$.
    Therefore, one can check that
        \begin{equation*}
           (x_1,\ldots,x_{j-t},y_{j-t+1},\ldots,y_{j-t+s},x_{j+1},\ldots,x_n) = (y_1,\ldots,y_{j-t+s},x_{j+1},\ldots,x_{j+s},y_{j+s+1},\ldots,y_n).
        \end{equation*}
    Thus, $B_{t,s}(\boldsymbol x)\cap B_{t,s}(\boldsymbol y)\neq\varnothing$,
    which is again a contradiction to the hypothesis that $\mathcal{C}$ is a $(t,s)$-burst correcting code.
  \end{enumerate}
\end{IEEEproof}

\begin{remark}
Note that in \cite{Levenshtein-2Burst-1970} Levenshtein proved a stronger result that $t$-deletion-correcting codes and $t$-insertion-correcting codes are also equivalent to codes correcting arbitrary $t_1$ deletions and $t_2$ insertions, as long as $t_1+t_2=t$.
It is tempting to ask if a similar result holds for the burst error model:
Is there a general equivalence between $(t_1,s_1)$-burst correcting codes and $(t_2,s_2)$-burst correcting codes, whenever $t_1+s_1=t_2+s_2$?
However, this is not true. For example, consider the following three strings $\boldsymbol x = \boldsymbol u 00100 \boldsymbol v$, $\boldsymbol y = \boldsymbol u 11111 \boldsymbol v$, and $\boldsymbol z = \boldsymbol u 01010 \boldsymbol v$.
$\boldsymbol x$ and $\boldsymbol y$ may belong to a $(3,1)$-burst correcting code, but they cannot belong to a $(2,2)$-burst correcting code,
since $\boldsymbol u 11100 \boldsymbol v\in B_{2,2}(\boldsymbol x)\cap B_{2,2}(\boldsymbol y)$.
$\boldsymbol y$ and $\boldsymbol z$ may belong to a $(2,2)$-burst correcting code, but they cannot belong to a $(3,1)$-burst correcting code,
since $\boldsymbol u 011 \boldsymbol v\in B_{3,1}(\boldsymbol y)\cap B_{3,1}(\boldsymbol z)$.
Thus $(3,1)$-burst correcting codes and $(2,2)$-burst correcting codes are not equivalent.
\end{remark}

\section{An Upper Bound on The Code Size}\label{sec:upper}

There are several upper bounds on the size of burst-deletion-correcting codes. Levenshtein\cite{Levenshtein-2Burst-1970} provided an asymptotic upper bound on the size of $t$-burst-deletion-correcting codes: if $\mathcal{C}\subseteq\mathbb{F}_2^n$ is a $t$-burst-correcting code, then $\vert\mathcal{C}\vert\leq 2^{n-t+1}/n$, which implies that the redundancy is asymptotically at least $\log n+t-1$. Schoeny \textit{et al.}\cite{Schoeny-Burst-IT17} constructed a hypergraph whose vertices are all sequences of $\mathbb{F}_2^{n-t}$ and the hyperedges are the $t$-burst-deletion balls of all sequences in $\mathbb{F}_2^n$. In this way the problem turns into analyzing the matching number of the hypergraph and they provided an explicit upper bound on the size of $t$-burst-deletion-correcting codes as $\vert\mathcal{C}\vert\leq (2^{n-t+1}-2^t)/(n-2t+1)$.

A sphere-packing type upper bound is usually not easy to obtain for most models concerning insertions and deletions, since the size of the corresponding error ball usually depends on the choice of the center. However, in this section we derive an unexpected result that the size of a $(t,s)$-burst ball is in fact a constant independent of the center, and thus give a sphere-packing type upper bound on the size of $(t,s)$-burst correcting codes. To analyze the size of the error ball, we first claim that $B_{t,s}(\boldsymbol x)$ can be partitioned into several disjoint parts.

For a $(t,s)$-burst of $\boldsymbol x$, let the deleted symbols be $x_i,\ldots,x_{i+t-1}$ where $1\leq i\leq n-t+1$, and the inserted symbols be $y_1,y_2,\ldots,y_s$. Observe that, for example, if the deleted symbols are 10101 and the inserted symbols are 111, then essentially this $(5,3)$-burst can be seen as a $(3,1)$-burst deleting 010 and inserting 1. Following this observation, let $B'_{k,\ell}(\boldsymbol x)$ denote the sequences obtained by a $(k,\ell)$-burst over $\boldsymbol x$, where the first and last inserted symbols are different from the first and last deleted symbols, respectively. In other words,
  \begin{equation*}
   B'_{k,\ell}(\boldsymbol x)= \bigcup_{i=1}^{n-k+1}\Big\{x_1\ldots x_{i-1}y_1\ldots y_{\ell}x_{i+k}\ldots x_n,
\text{ where }x_i\neq y_1, x_{i+k-1}\neq y_\ell \Big\}.
  \end{equation*}
For example, when $k=1$ and $\ell\geq 1$, we have $y_1=y_{\ell}\neq x_i$, i.e., we delete a symbol $a\in\{0,1\}$ and insert a sequence of length $\ell$ starting and ending with $1-a$. Similarly, when $\ell=1$ and $k\geq 1$, we have $x_i=x_{i+k-1}\neq y_1$, i.e., we delete a sequence of length $k$ starting and ending with the symbol $a$ and insert $1-a$. When $k=0$ or $\ell=0$, $B'_{0,\ell}$ or $B'_{k,0}$ is simply the burst insertion ball or the burst deletion ball. In particular, $B'_{0,0}=\{\boldsymbol x\}$ contains the sequence $\boldsymbol x$ itself.

\begin{lemma}\label{ballpar1}
When $t\geq s$,  $B_{t,s}(\boldsymbol x)$ is a disjoint union of $B'_{t-s+\ell,\ell}(\boldsymbol x)$, $0\leq \ell \leq s$. Similarly, when $t\leq s$, $B_{t,s}(\boldsymbol x)$ is a disjoint union of $B'_{k,s-t+k}(\boldsymbol x)$, $0\leq k \leq t$.
\end{lemma}
\begin{IEEEproof}
We only prove the first part. The second part can be proved analogously.

First we need to prove that $\bigcup_{\ell=0}^{s}B'_{t-s+\ell,\ell}(\boldsymbol x) = B_{t,s}(\boldsymbol x)$. Obviously, $\bigcup_{\ell=0}^{s}B'_{t-s+\ell,\ell}(\boldsymbol x)\subseteq B_{t,s}(\boldsymbol x)$. For the other direction, consider
$\boldsymbol w\in  B_{t,s}(\boldsymbol x)$ and suppose $\boldsymbol w$ is obtained by deleting $x_i,\ldots,x_{i+t-1}$ from $\boldsymbol x$ and inserting $y_1,\ldots,y_s$. Let $p\in\{1,2,\dots,s\}$ be the smallest index such that $y_p\neq x_{i+p-1}$, and let $q\in\{1,2,\dots,s\}$ be the largest index such that $y_q\neq x_{i+t+q-s-1}$.
\begin{itemize}
  \item If such $p$ and $q$ exist and $p\leq q$, then
$\boldsymbol w$ can be seen as deleting $x_{i+p-1},\ldots,x_{i+t+q-s-1}$ and inserting $y_p,\ldots,y_q$. Thus $\boldsymbol w\in
B'_{t-s+q-p+1,q-p+1}(\boldsymbol x)$, where $q-p+1\in\{1,\ldots,s\}$.
  \item If $p$ does not exist, it means that $(y_1,\ldots,y_s)=(x_i,\ldots,x_{i+s-1})$, i.e., the inserted sequence is a prefix of the deleted sequence. Then $\boldsymbol w$ can be seen as deleting $(x_{i+s},\ldots,x_{i+t-1})$ and thus $\boldsymbol w\in B'_{t-s,0}(\boldsymbol x)$.
  \item If $q$ does not exist, it means that $(y_1,\ldots,y_s)=(x_{i+t-s},\ldots,x_{i+t-1})$, i.e., the inserted sequence is a suffix of the deleted sequence. Then $\boldsymbol w$ can be seen as deleting $(x_{i},\ldots,x_{i+t-s-1})$ and thus $\boldsymbol w\in B'_{t-s,0}(\boldsymbol x)$.
  \item Finally, if $p$ and $q$ exist and $p>q$, then it means that $(y_1,\ldots,y_{p-1})$ is a prefix of the deleted sequence and $(y_{q+1},\ldots,y_s)$ is a suffix of the deleted sequence. Then $\boldsymbol w$ can be also seen as deleting $(x_{i+p-1},\ldots,x_{i+p+t-s-2})$ and thus $\boldsymbol w\in B'_{t-s,0}(\boldsymbol x)$.
\end{itemize}

Therefore, $\bigcup_{\ell=0}^{s}B'_{t-s+\ell,\ell}(\boldsymbol x) = B_{t,s}(\boldsymbol x)$ holds and it only remains to show the disjointness of $B'_{t-s+\ell,\ell}(\boldsymbol x)$ for $0\leq \ell \leq s$. Suppose there exist a sequence $\boldsymbol z\in B'_{t-s+\ell_1,\ell_1}(\boldsymbol x)\cap B'_{t-s+\ell_2,\ell_2}(\boldsymbol x)$, and $\boldsymbol z$ is obtained by deleting $t-s+\ell_1$ (resp. $t-s+\ell_2$) symbols and inserting $\ell_1$ (resp. $\ell_2$) symbols at the $i_1$th (resp. $i_2$th) coordinate of $\boldsymbol x$.

If $i_1=i_2\triangleq i$, without loss of generality, we can assume $\ell_1<\ell_2$. Then $\boldsymbol z$ has the following two representations:
  \begin{eqnarray*}
    & \boldsymbol z^1= (x_1,\ldots,x_{i-1},y_1^1,\ldots,y_{\ell_1}^1,x_{i+t-s+\ell_1+1},\ldots,x_n), \\
    & \boldsymbol z^2=(x_1,\ldots,x_{i-1},y_1^2,\ldots,y_{\ell_2}^2,x_{i+t-s+\ell_2+1},\ldots,x_n).
  \end{eqnarray*}

Comparing the $(i+t-s+\ell_2)$th symbol, we have $x_{i+t-s+\ell_2}=y_{\ell_2}^2$. However, since $\boldsymbol z\in B'_{t-s+\ell_2,\ell_2}(\boldsymbol x)$,  in the representation of $\boldsymbol z^2$ the last deleted symbol $x_{i+t-s+\ell_2}$ should be different from the last inserted symbol $y_{\ell_2}^2$, a contradiction.

If $i_1\neq i_2$, without loss of generality let $i_1<i_2$. Comparing the $i_1$th symbol of $\boldsymbol z^1$ and $\boldsymbol z^2$, we have $y_1^1=x_{i_1}$. However, since $\boldsymbol z\in B'_{t-s+\ell_1,\ell_1}(\boldsymbol x)$, in the representation of $\boldsymbol z^1$ the first deleted symbol $x_{i_1}$ should be different from the first inserted symbol $y_1^1$, again a contradiction.

Thus, $B'_{t-s+\ell_1,\ell_1}(\boldsymbol x)$ and $B'_{t-s+\ell_2,\ell_2}(\boldsymbol x)$ are disjoint for any $\ell_1\neq \ell_2$, and the lemma is proved.
\end{IEEEproof}

Next, we need to calculate the size of each $B'_{k,\ell}(\boldsymbol x)$.

When $\ell=0$, in \cite{Levenshtein-2Burst-1970} Levenshtein calculated the size of a $k$-burst-deletion ball as:
\begin{equation}\label{Equ:t-burst-ball}
  \vert B'_{k,0}(\boldsymbol x)\vert=1+\sum_{i=1}^{k}(r(A_k(\boldsymbol x)_i)-1),
\end{equation}
where $k\geq1$, $A_k(\boldsymbol x)_i$ is the $i$th row of the array representation $A_k(x)$, and $r(A_k(\boldsymbol x)_i)$ is the number of runs in the $i$th row. Further note that for the case $k=\ell=0$ we have $B'_{0,0}(\boldsymbol x)=\{\boldsymbol x\}$ and thus $\vert B'_{0,0}(\boldsymbol x) \vert=1$, which can be also seen as a special case of Equation (\ref{Equ:t-burst-ball}).
In the next lemmas, we will compute the size of $ B'_{k,\ell}(\boldsymbol x)$ for the remaining cases.

\begin{lemma}\label{ballcal1}
  Let $\boldsymbol x\in\mathbb{F}_2^n$, $k\geq 1,\ell=1$, then
\begin{equation}\label{Equ:k-1-burst}
\vert B'_{k,1}(\boldsymbol x)\vert=n-\sum_{i=1}^{k-1}r(A_{k-1}(\boldsymbol x)_i).
\end{equation}
\end{lemma}
\begin{IEEEproof}
First consider $k\geq 2$. Recall that any $\boldsymbol y \in B'_{k,1}(\boldsymbol x)$ is obtained from $\boldsymbol x$ by deleting a string of length $k$ starting and ending with the same symbol $a$ and inserting $1-a$. Let $\boldsymbol y_1,\boldsymbol y_2\in B'_{k,1}(\boldsymbol x)$ where the $(k,1)$-bursts start at the $i_1$th and $i_2$th coordinate, respectively (WLOG, assume $i_1<i_2$). That is, $\boldsymbol y_1=(x_1,\dots,x_{i_1-1},y_1,x_{i_1+k},\dots,x_n)$ and $\boldsymbol y_2=(x_1,\dots,x_{i_2-1},y_2,x_{i_2+k},\dots,x_n)$. Note that since $x_{i_1}=x_{i_1+k-1}\neq y_1$, then $\boldsymbol y_1$ and $\boldsymbol y_2$ have distinct symbols on their $i_1$th coordinate and thus $\boldsymbol y_1 \neq \boldsymbol y_2$. Therefore, $\vert B'_{k,1}(\boldsymbol x)\vert$ is exactly the number of coordinates $i$ such that $x_i=x_{i-k+1}$, $1\leq i \leq n-k+1$.

Write $\boldsymbol x$ as a $(k-1)\times\frac{n}{k-1}$ array:
\begin{equation*}
        A_{k-1}(\boldsymbol x)=\begin{bmatrix}
        x_1 & x_{k} & \cdots & x_{n-k+2} \\
        x_2 & x_{k+1} & \cdots & x_{n-k+3} \\
        \vdots & \vdots & \ddots & \vdots \\
        x_{k-1} & x_{2k-2} & \cdots & x_n
      \end{bmatrix}.
\end{equation*}
In the representation of $A_{k-1}(\boldsymbol x)$, $x_i$ and $x_{i-k+1}$ are two consecutive symbols in a row. The $i$th row contributes $\frac{n}{k-1}-r(A_{k-1}(\boldsymbol x)_i)$ to the size of $\vert B'_{k,1}(\boldsymbol x)\vert$. Therefore for $k\geq 2$ we have
\begin{equation*}
     \vert B'_{k,1}(\boldsymbol x)\vert = \sum_{i=1}^{k-1}\Big(\frac{n}{k-1}-r(A_{k-1}(\boldsymbol x)_i)\Big)
      = n-\sum_{i=1}^{k-1}r(A_{k-1}(\boldsymbol x)_i).
\end{equation*}

In addition, when $k=1$, $B'_{1,1}(\boldsymbol x)$ is the Hamming sphere of radius 1 (without the center) and thus $\vert B'_{1,1}(\boldsymbol x) \vert =n$, which can also be seen as a special case of Equation (\ref{Equ:k-1-burst}).
\end{IEEEproof}

\begin{lemma}\label{ballcal2}
  Let $\boldsymbol x\in\mathbb{F}_2^n$, $k\geq 1,\ell\geq2$, then
  $$\vert B'_{k,\ell}(\boldsymbol x)\vert=(n-k+1)\cdot2^{\ell-2}.$$
\end{lemma}
\begin{IEEEproof}
All the sequences in $B'_{k,\ell}(\boldsymbol x)$ have the form $x_1\ldots x_{i-1}y_1\ldots y_{\ell}x_{i+k}\ldots x_n$, where $1\leq i\leq n-k+1$ is the starting coordinate of the deletion. We claim that if the starting coordinates are different, then the two obtained sequence must be different. Suppose $\boldsymbol y_1$ and $\boldsymbol y_2$ are two sequences in $B'_{k,\ell}(\boldsymbol x)$, and their errors start at the $i_1$th and $i_2$th coordinate of $\boldsymbol x$, respectively (WLOG, assume $i_1< i_2$). Then $\boldsymbol y_1$ and $\boldsymbol y_2$ have distinct symbols on their $i_1$th coordinate and thus $\boldsymbol y_1\neq \boldsymbol y_2$.

To calculate $\vert B'_{k,\ell}(\boldsymbol x)\vert$, we have $n-k+1$ choices for $i$, and once $i$ is given the inserted sequence has $2^{\ell-2}$ possibilities since $y_1=1-x_i$ and $y_{\ell}=1-x_{i+k-1}$ are fixed. As a consequence, there are altogether $(n-k+1)\cdot2^{\ell-2}$ distinct sequences in $ B'_{k,\ell}(\boldsymbol x)$.
\end{IEEEproof}

\begin{lemma}\label{ballcal3}
Let $\boldsymbol x\in\mathbb{F}_2^n$, $k=0,\ell\geq1$, then
  $$\vert B'_{0,\ell}(\boldsymbol x)\vert=n\cdot2^{\ell-1}+2^{\ell}.$$
\end{lemma}
\begin{IEEEproof}
For any $\boldsymbol y \in B'_{0,\ell}(\boldsymbol x)$, let $1\leq i\leq n$ be the first coordinate where $\boldsymbol x$ and $\boldsymbol y$ differ, then $\boldsymbol y$ can be seen as inserting a sequence of length $\ell$ at the $i$th coordinate and there are $2^{\ell-1}$ possibilities for the inserted sequence.
If no such $i$ exists, which means the first $n$ symbols of $\boldsymbol y$ equal $\boldsymbol x$, then $\boldsymbol y$ can be seen as inserting a sequence of length $\ell$ at the end of $\boldsymbol x$ and there are $2^{\ell}$ possibilities for the inserted sequence. To sum up we have $\vert B'_{0,\ell}(\boldsymbol x)\vert=n\cdot2^{\ell-1}+2^{\ell}$.
\end{IEEEproof}

Summarizing the previous lemmas, the size of $\vert B'_{k,\ell}(\boldsymbol x)\vert$ is as follows:
  \begin{equation*}
    B'_{k,\ell}(\boldsymbol x)=\begin{cases}
                              1+\sum_{i=1}^{k}(r(A_k(\boldsymbol x)_i)-1), & \mbox{if } k\geq0,\ell=0 \\
                              n-\sum_{i=1}^{k-1}r(A_{k-1}(\boldsymbol x)_i), & \mbox{if } k\geq1,\ell=1 \\
                              (n-k+1)\cdot2^{\ell-2}, & \mbox{if } k\geq1,\ell\geq2 \\
                              n\cdot2^{\ell-1}+2^\ell, & \mbox{if } k=0,\ell\geq1
                            \end{cases}
  \end{equation*}

Now we are ready to compute the size of the error ball $B_{t,s}(\boldsymbol x)$.

\begin{theorem}\label{theorem2}
For $\boldsymbol x\in\{0,1\}^n,t,s\in\mathbb{N}^*$, the size of a $(t,s)$-burst error ball is $$\vert B_{t,s}(\boldsymbol x)\vert=(n-t+2)\cdot 2^{s-1}.$$
\end{theorem}

\begin{IEEEproof}
If $t\geq s$, from Lemma \ref{ballpar1}, we can calculate
\begin{align*}
  \vert B_{t,s}(\boldsymbol x)\vert & =\sum_{\ell=0}^{s}\vert B'_{t-s+\ell,\ell}(\boldsymbol x)\vert \\
   & =1+\sum_{i=1}^{t-s}(r(A_{t-s}(\boldsymbol x)_i)-1)+n-\sum_{i=1}^{t-s}r(A_{t-s}(\boldsymbol x)_i)+\sum_{\ell=2}^{s}(n-t+s-\ell+1)\cdot2^{\ell-2} \\
   & =n+s-t+1+\sum_{\ell=0}^{s-2}(n-t+s-\ell-1)\cdot2^\ell
\end{align*}

Let $S=\sum_{\ell=0}^{s-2}(n-t+s-\ell-1)\cdot2^\ell$, then
 $2S=\sum_{\ell=0}^{s-2}(n-t+s-\ell-1)\cdot2^{\ell+1}=\sum_{\ell=1}^{s-1}(n-t+s-\ell)\cdot2^{\ell}$.
So, we can calculate $S$ as
\begin{align*}
  2S-S & =\sum_{\ell=1}^{s-1}(n-t+s-\ell)\cdot2^{\ell}-\sum_{\ell=0}^{s-2}(n-t+s-\ell-1)\cdot2^\ell \\
   & =(n-t+1)\cdot2^{s-1}+\sum_{\ell=1}^{s-2}2^\ell-(n-t+s-1) \\
   & =(n-t+2)\cdot2^{s-1}-n+t-s-1.
\end{align*}

Thus, the size of the error ball is
\begin{align*}
  \vert B_{t,s}(\boldsymbol x)\vert & =n+s-t+1+\sum_{\ell=0}^{s-2}(n-t+s-i-1)\cdot2^\ell \\
  & =n+s-t+1+(n-t+2)\cdot2^{s-1}-n+t-s-1 \\
  & =(n-t+2)\cdot2^{s-1}.
\end{align*}

Similarly, if $t<s$, from Lemma \ref{ballpar1}, we can calculate
\begin{align*}
  \vert B_{t,s}(\boldsymbol x)\vert & =\sum_{k=0}^{t}\vert B'_{k,s-t+k}(\boldsymbol x)\vert \\
   & = n\cdot2^{s-t-1}+2^{s-t}+\sum_{k=1}^{t}(n-k+1)\cdot2^{s-t+k-2}\\
   & = (n-t+2)\cdot2^{s-1}.
\end{align*}


Summing up the above, we have proved that for any $t,s\in\mathbb{N}^*$, $\vert B_{t,s}(\boldsymbol x)\vert=(n-t+2)\cdot2^{s-1}$.
\end{IEEEproof}

\begin{example}
  Let $n=12, \boldsymbol x=101011100100$, $t=4,s=1$, then
  \begin{equation*}
        A_{3}(\boldsymbol x)=\begin{bmatrix}
        1 & 0 & 1 & 1 \\
        0 & 1 & 0 & 0 \\
        1 & 1 & 0 & 0
      \end{bmatrix}.
\end{equation*}
\begin{equation*}
\begin{split}
 B'_{3,0}(\boldsymbol x)=\{& 011100100,111100100,101100100,101000100,101010100,101011100\}, \\
 B'_{4,1}(\boldsymbol x)=\{& 100100100,101011000,101011110,101011101\}.
\end{split}
\end{equation*}
Hence, $\vert B'_{3,0}(\boldsymbol x)\vert +\vert B'_{4,1}(\boldsymbol x)\vert=10=(12-4+2)\cdot2^0.$
\end{example}

Theorem \ref{theorem2} shows that the size of the $(t,s)$-burst ball is independent of its center, which is rather rare in other models concerning deletions or insertions. A sphere-packing type upper bound naturally follows.

\begin{theorem}\label{theorem3}
  Let $\mathcal{C}\subseteq\mathbb{F}_2^n$ be a $(t,s)$-burst correcting code, then
  $$\vert\mathcal{C}\vert\leq\frac{2^{n-t+1}}{n-t+2}.$$
\end{theorem}
\begin{IEEEproof}
Given any two words $\boldsymbol c_1,\boldsymbol c_2\in\mathcal{C}$,
$$B_{t,s}(\boldsymbol c_1)\cap B_{t,s}(\boldsymbol c_2)=\varnothing.$$
Now consider the union of all $(t,s)$-burst balls centered at the codewords in $\mathcal{C}$. Obviously, their union is a subset of $\mathbb{F}_2^{n-t+s}$. In other words,
$$\Big\vert\bigcup_{\boldsymbol c\in\mathcal{C}}B_{t,s}(\boldsymbol c)\Big\vert\leq 2^{n-t+s}.$$
Since any two $(t,s)$-burst balls centered at distinct codewords in $\mathcal{C}$ are disjoint,
\begin{equation*}
     \Big\vert\bigcup_{\boldsymbol c\in\mathcal{C}}B_{t,s}(\boldsymbol c)\Big\vert  =\sum_{\boldsymbol c\in\mathcal{C}}\vert B_{t,s}(\boldsymbol c)\vert
       =(n-t+2)\cdot2^{s-1}\cdot\vert\mathcal{C}\vert
\end{equation*}
Consequently, $$\vert\mathcal{C}\vert\leq\frac{2^{n-t+1}}{n-t+2}.$$
\end{IEEEproof}

Note that for $t=s=1$ the upper bound above coincides with the Hamming bound for a single substitution, and when $t=s$ our upper bound coincides with the upper bound of burst-error correcting code for substitutions derived by Dass\cite{Dass-Subburst-JIOC80}.
Moreover, it is quite interesting that the upper bound above is irrelevant to the number of inserted symbols $s$.
Recall the equivalence of $(t,s)$-burst correcting codes and $(s,t)$-burst correcting codes, $t$ can be replaced by $s$ in the upper bound of Theorem \ref{theorem3}, and the better bound is the one using the relatively larger value between $t$ and $s$. Therefore, from now on we only focus on the case $t\geq s$, and the redundancy of a $(t,s)$-burst correcting code is lower bounded by $$n-\log|\mathcal{C}| \geq \log (n-t+2)+t-1\approx \log n+t-1.$$

\section{A General Construction of $(t,s)$-Burst Correcting Codes for $t\geq2s$}\label{sec:t-s-constr}

The equivalence allows us to focus only on the case $t\geq s$, i.e., the number of deletions is no less than the number of insertions. In this section, we provide a construction of $(t,s)$-burst correcting codes for $t\geq2s$. We start with an overall explanation of the main idea, and then proceed with the detailed construction and analysis followed by some discussions.

\subsection{Sketch of the main Idea}
In the array representation $A_{t-s}(\boldsymbol x)$,
if $t\geq2s$, then it is routine to check that a $(t,s)$-burst results in an array of size $(t-s)\times (\frac{n}{t-s}-1)$ and each row of $A_{t-s}(\boldsymbol x)$ suffers from either one deletion or a $(2,1)$-burst. To be more specific, suppose the starting coordinate of the error is located at the $k$th row of $A_{t-s}(\boldsymbol x)$, then each of the rows indexed by $k,k+1,\dots,k+s-1$ (indices are calculated modulo $t-s$) suffers from a $(2,1)$-burst and each of the rest rows suffers from a single deletion. The following toy example help visualize the error type on each row. Let $t=5$, $s=2$, and $n=12$. Below are the error patterns if the error starts on the 3rd, 4th, or the 5th coordinate, correspondingly.
\begin{equation*}
    \begin{bmatrix}
     x_1 & \underline{x_4} & \underline{x_7} & x_{10} \\
     x_2 & \underline{x_5} & x_8 & x_{11} \\
     \underline{x_3} & \underline{x_6} & x_9 & x_{12}
                     \end{bmatrix}
                     \rightarrow
    \begin{bmatrix}
     x_1 & \underline{y_2} & x_{10} \\
     x_2 & x_8 & x_{11} \\
     \underline{y_1} & x_9 & x_{12}
     \end{bmatrix},
    \begin{bmatrix}
     x_1 & \underline{x_4} & \underline{x_7} & x_{10} \\
     x_2 & \underline{x_5} & \underline{x_8} & x_{11} \\
     x_3 & \underline{x_6} & x_9 & x_{12}
                     \end{bmatrix}
                     \rightarrow
    \begin{bmatrix}
     x_1 & \underline{y_1} & x_{10} \\
     x_2 & \underline{y_2} & x_{11} \\
     x_3 & x_9 & x_{12}
     \end{bmatrix},
     \begin{bmatrix}
     x_1 & x_4 & \underline{x_7} & x_{10} \\
     x_2 & \underline{x_5} & \underline{x_8} & x_{11} \\
     x_3 & \underline{x_6} & \underline{x_9} & x_{12}
                     \end{bmatrix}
                     \rightarrow
    \begin{bmatrix}
     x_1 & x_4 & x_{10} \\
     x_2 & \underline{y_1} & x_{11} \\
     x_3 & \underline{y_2} & x_{12}
     \end{bmatrix}
\end{equation*}

Recall that the $(2,1)$-burst correcting code as shown in Equation (\ref{Equ:2-1-burst}) can also correct a single deletion, due to the VT syndrome in its definition. We can let the first row of $A_{t-s}(\boldsymbol x)$ be chosen from a $(2,1)$-burst correcting code, and decoding this row will provide us some additional knowledge about the location of the error in the other rows. To be more specific, the decoding of the first row (using a decoder for a $(2,1)$-burst correcting code) falls into three cases.

\begin{itemize}
  \item[1.] The first row suffers from a $(2,1)$-burst where the error is of the form $00\rightarrow 1$ or $11\rightarrow 0$.
  \item[2.] The first row suffers from a $(2,1)$-burst which does not belong to Case 1.
  \item[3.] The first row suffers from a single deletion.
\end{itemize}

For Case 1, we can uniquely determine the two erroneous coordinates in the first row. Since $t\leq 2(t-s)$, the deleted coordinates are within at most three consecutive columns. Therefore, the erroneous coordinates in the other rows are located within an interval of length at most 3. See the previous example, if we have determined that the first row suffers from a $(2,1)$-burst at the second coordinate (the left one and the middle one), then the erroneous coordinates on the other rows are only of the following possibilities: a $(2,1)$-burst at the first coordinate (3rd row in the left one), a $(2,1)$-burst at the second coordinate (2nd row in the middle one), or a single deletion at the second coordinate.

Case 2 can be also seen as a single deletion in the first row. Therefore we can combine Case 2 and Case 3 together, and locate the single deletion in the first row within a run, which might range from the $c_1$th to the $c_2$th column, for some $c_1\le c_2$. Then the errors in the other rows should be within columns indexed from $c_1-1$ to $c_2$. By adding additional {\it run-length constraint} on the first row, $c_2-c_1$ is bounded and then we may apply a \textit{$(2,1)$-burst-SVT code} (to be defined later), which can correct a $(2,1)$-burst with the additional knowledge about the location of the error within an interval of $P$ consecutive coordinates.

Here is an example to illustrate the whole decoding process.

\begin{example}
  Let $\boldsymbol x=101011001101110$, $t=4,s=1$, and the erroneous sequence is $\boldsymbol x'=101010101110$. First write $\boldsymbol x'$ in the form of an array,
  \begin{equation*}
    A_3(\boldsymbol x')=\begin{bmatrix}
                          1 & 0 & 1 & 1 \\
                          0 & 1 & 0 & 1 \\
                          1 & 0 & 1 & 0
                        \end{bmatrix}.
  \end{equation*}

Applying the decoder of a $(2,1)$-burst correcting code to the first row, we get $10011$. In addition we know that the first row suffers from a single deletion on either the second or the third coordinate. Therefore, the errors in the remaining two rows are within their first three coordinates. Applying the decoder of a $(2,1)$-burst SVT code with $P=3$ to them, we get the second and third rows as $01001$, $11110$ respectively. Consequently, \begin{equation*}
    A_3(\boldsymbol x)=\begin{bmatrix}
                          1 & 0 & 0 & 1 & 1 \\
                          0 & 1 & 0 & 0 & 1 \\
                          1 & 1 & 1 & 1 & 0
                        \end{bmatrix},
  \end{equation*}
  and $\boldsymbol x=101011001101110$ is correctly decoded.
\end{example}

We close this subsection by discussing the parameter $P$ in the run-length constraint\cite{Immink-coding-1991}. If $P$ is too large (e.g. linear in $n$), using $(2,1)$-burst SVT codes does not reduce the total redundancy as compared to just using $(2,1)$-burst correcting codes. Therefore, to further reduce the redundancy we want to make $P$ small. Define $\mathcal{S}_n(f(n))$ to be the set of binary sequences of length $n$ whose maximum run length is at most $f(n)$. An algorithm is provided to efficiently encode any binary sequence to a $(\log n+3)$-RLL sequence with only 1 bit of redundancy in \cite{Schoeny-Burst-IT17}, i.e., $\vert\mathcal{S}_n(\log n+3)\vert\geq 2^{n-1}$. Once we use a subcode of $\mathcal{S}_n(\log n+3)$ in the first row, it is guaranteed that for the other rows we have the additional knowledge about the location of the error within a bounded interval.

Now, we are fully prepared to present our main construction.

\subsection{A general construction for $t\geq2s$}
We follow the array representation framework to write $\boldsymbol x$ as an array $A_{t-s}(\boldsymbol x)$ of size $(t-s)\times \frac{n}{t-s}$. The first row $A_{t-s}(\boldsymbol x)$ comes from the following code.

\begin{construction}
For arbitrary integers $n$ and $a\in\mathbb{Z}_{2n-1}, b\in\mathbb{Z}_4$, define  $\mathcal{C}_{2,1}^{\mathrm{RLL}}(n;a,b,\log n+3)$ as
$$\mathcal{C}_{2,1}^{\mathrm{RLL}}(n;a,b,\log n+3)=\mathcal{C}_{2,1}(n;a,b)\cap\mathcal{S}_n(\log n+3).$$
\end{construction}

\begin{theorem}\label{theo21RLL}
For arbitrary integers $n$, there exists a code $\mathcal{C}_{2,1}^{\mathrm{RLL}}(n;a,b,\log n+3)$ with redundancy at most $\log n+4$.
\end{theorem}
\begin{IEEEproof}
Recall that $\vert\mathcal{S}_n(\log n+3)\vert\geq 2^{n-1}$. Moreover, for $a\in\mathbb{Z}_{2n-1}, b\in\mathbb{Z}_4$, $\bigcup_{a,b} \mathcal{C}_{2,1}(n;a,b)\cap\mathcal{S}_n(\log n+3)$ is a disjoint partition of $\mathcal{S}_n(\log n+3)$. Thus according to the pigeonhole principle, there must exist choices for $a\in\mathbb{Z}_{2n-1}$ and $b\in\mathbb{Z}_4$, such that
$$\vert\mathcal{C}_{2,1}^{\mathrm{RLL}}(n;a,b,\log n+3)\vert\geq \frac{2^{n-1}}{4(2n-1)}.$$
Therefore, the redundancy is at most $$n-\log\vert\mathcal{C}_{2,1}^{\mathrm{RLL}}(n;a,b,\log n+3)\vert = \log4(2n-1)+1<\log n+4.$$
\end{IEEEproof}

Due to the RLL constraint of the first row, the starting coordinate of the error on each remaining row will be limited to an interval of length $\log n+4$. Now, we will provide a code which can correct a $(2,1)$-burst with this additional knowledge.

\begin{construction}
  For arbitrary integers $n$ and $c\in\mathbb{Z}_{2P-1}, d\in\mathbb{Z}_4$, $P\leq n$ define the $SVT_{2,1}^{\mathrm{burst}}(n;c,d,P)$ as
  \begin{equation*}
      SVT_{2,1}^{\mathrm{burst}}(n;c,d,P)=\Big\{\boldsymbol x:  VT(\boldsymbol x)\equiv c\pmod{2P-1},
        \sum_{i=1}^{n}x_i\equiv d\pmod{4}\Big\}.
  \end{equation*}
\end{construction}


\begin{theorem}
The $(2,1)$-burst-SVT code $SVT_{2,1}^{\mathrm{burst}}(n;c,d,P)$ can correct a $(2,1)$-burst with the additional knowledge of the starting location of the $(2,1)$-burst within an interval of $P$ consecutive coordinates. Furthermore, there exist choices for $c$ and $d$ such that the redundancy of the code is at most $\log P+3$.
\end{theorem}

\begin{IEEEproof}
For any $\boldsymbol u\in\mathbb{F}_2^n$ that suffers from a $(2,1)$-burst, denote the received sequence as $\boldsymbol u'\in\mathbb{F}_2^{n-1}$. Define $\Delta=\sum_{i=1}^{n}u_i-\sum_{i=1}^{n-1}u'_i\pmod4$, $\Delta\in\{0,1,2,3\}$. The decoding starts with the observation of the value $\Delta$.

We have $\Delta=3$ if and only if $00\rightarrow1$ happens. Similarly, We have $\Delta=2$ if and only if $11\rightarrow0$ happens. For the remaining cases, when $\Delta=0$ or $1$, the $(2,1)$-burst error could be seen as just a single deletion. Since $SVT_{2,1}^{\mathrm{burst}}(n;c,d,P)$ is also an SVT code (due to the VT syndrome in its definition), it can correct a single deletion with the additional knowledge of the deleted coordinate within a length-$P$ interval.

We are only left with the case when the error is $11\rightarrow0$ or $00\rightarrow1$. We only prove the former case $11\rightarrow0$. The latter case $00\rightarrow1$ can be proved analogously.

Suppose there exist $\boldsymbol x, \boldsymbol y\in SVT_{2,1}^{\mathrm{burst}}(n;c,d,P)$, such that $\boldsymbol z \in B_{2,1}(\boldsymbol x)\cap B_{2,1}(\boldsymbol y)$.
Suppose the $(2,1)$-burst of the form $11\rightarrow0$ starts at $i$th coordinate of $\boldsymbol x$ and $j$th coordinate of $\boldsymbol y$, and without loss of generality $i<j$.
We have
\begin{equation*}
      \begin{split}
        &\boldsymbol x=(x_{1},\ldots,x_{i-1},1,1,x_{i+2},\ldots,x_{j},0,x_{j+2},\ldots,x_{n}), \\ &\boldsymbol y=(y_{1},\ldots,y_{i-1},0,y_{i+1},\ldots,y_{j-1},1,1,y_{j+2},\ldots,y_{n}),
      \end{split}
\end{equation*}
where $x_k=y_k$ when $1\leq k\leq i-1$ and $j+2\leq k\leq n$, $x_{k+1}=y_k$ when $i+1\leq k\leq j-1$.

Now we consider the difference of syndromes $VT(\boldsymbol x)-VT(\boldsymbol y)$, which is equal to
\begin{equation*}
\sum_{i=1}^{n}ix_i-\sum_{i=1}^{n}iy_i
 = i+(i+1)+\text{wt}(x_{i+2},\ldots,x_{j})-j-(j+1)
 = 2(i-j)+\text{wt}(x_{i+2},\ldots,x_{j})
\end{equation*}
Since $0\leq\text{wt}(x_{i+2},\ldots,x_{j})\leq j-i-1$, we have $$2(i-j)\leq VT(\boldsymbol x)-VT(\boldsymbol y)\leq i-j-1.$$
Furthermore, $i$ and $j$ are within an interval of length $P$, so $1\leq j-i\leq P-1$. Hence, $$-2(P-1)\leq VT(\boldsymbol x)-VT(\boldsymbol y)\leq -2.$$
Therefore, $VT(\boldsymbol x)-VT(\boldsymbol y)\neq 0$,
which contradicts to the fact that $\boldsymbol x,\boldsymbol y\in SVT_{2,1}^{\mathrm{burst}}(n;c,d,P)$. Thus $SVT_{2,1}^{\mathrm{burst}}(n;c,d,P)$ can uniquely correct a $(2,1)$-burst error with the additional knowledge of the starting location of the $(2,1)$-burst within an interval of $P$ consecutive coordinates.

Moreover, since $\bigcup_{c,d} SVT_{2,1}^{\mathrm{burst}}(n;c,d,P)$ is a partition of $\mathbb{F}_2^n$, according to the pigeonhole principle, there must exist $c$ and $d$ such that the code size is at least $\frac{2^n}{4(2P-1)}$, thus the redundancy of the code is at most $\log (2P-1)+2<\log P+3$.
\end{IEEEproof}

Now, we are ready to present our construction of $(t,s)$-burst correcting codes for $t\geq2s$.

\begin{construction}\label{constr2s}
Let $t,s\in\mathbb{N}^*$ such that $t\geq2s$. Let $a\in\mathbb{Z}_{2n/(t-s)-1},b\in\mathbb{Z}_4$, and $c_i\in\mathbb{Z}_{2P-1},d_i\in\mathbb{Z}_4$, where $2\leq i\leq t-s$, $P=\log\frac{n}{t-s}+4$. The code $\mathcal{C}_{t,s} $ is constructed as follows:
\begin{equation*}
  \begin{split}
   \mathcal{C}_{t,s} \triangleq&\Big\{\boldsymbol x:A_{t-s}(\boldsymbol x)_1\in \mathcal{C}_{2,1}^{\mathrm{RLL}}\Big(\frac{n}{t-s};a,b,\log\frac{n}{t-s}+3\Big), \\
    & A_{t-s}(\boldsymbol x)_i\in SVT_{2,1}^{\mathrm{burst}}\Big(\frac{n}{t-s};c_i,d_i,\log\frac{n}{t-s}+4\Big),
    \text{ for }2\leq i\leq t-s\Big\}.
  \end{split}
\end{equation*}
\end{construction}

\begin{theorem}
  The code $\mathcal{C}_{t,s} $ is a $(t,s)$-burst correcting code, and there exist choices for $a,b,c_i,d_i$ such that the redundancy of the code is at most $\log n+(t-s-1)\log\log n+O(1)$.
\end{theorem}
\begin{IEEEproof}
Suppose $\boldsymbol x\in \mathcal{C}_{t,s} $ suffers from a $(t,s)$-burst and the erroneous sequence is denoted as $\boldsymbol y$. Consider them in the array form $A_{t-s}(\boldsymbol x)$ and $A_{t-s}(\boldsymbol y)$. The $(t,s)$-burst will cause either a $(2,1)$-burst or a single deletion in the first row, which could be correctly decoded since the first row comes from a $(2,1)$-burst correcting code. After decoding $A_{t-s}(\boldsymbol x)_1$ and comparing with $A_{t-s}(\boldsymbol y)_1$, due to the run-length constraint of the first row, we will get some additional knowledge.

On one hand, if the first row suffers from a single deletion, then we may locate its deleted coordinate within an interval of length $\log\frac{n}{t-s}+3$. The starting coordinate of error in the other rows will be located within an interval of length $\log\frac{n}{t-s}+4$. Then the rest rows can be uniquely decoded due to the property of the $(2,1)$-burst SVT code.

On the other hand, if the first row suffers from a $(2,1)$-burst in the form of $00\rightarrow 1$ or $11\rightarrow 0$, then we may locate the erroneous coordinates on the first row and the erroneous coordinates on the other rows are within an interval of length 3, thus again the rest rows can be uniquely decoded due to the property of the $(2,1)$-burst SVT code.

To sum up, $\mathcal{C}_{t,s} $ is indeed a $(t,s)$-burst correcting code.

As for the size of the code, by the pigeonhole principle there must exist choices for $a,b,c_i,d_i$, $2\leq i\leq t-s$, such that
$$\vert\mathcal{C}_{t,s} \vert\geq\frac{2^{\frac{n}{t-s}-1}\cdot(2^{\frac{n}{t-s}})^{t-s-1}}{4\cdot(\frac{2n}{t-s}-1)\cdot\Big(4\cdot\big(2(\log\frac{n}{t-s}+4)-1\big)\Big)^{t-s-1}}.$$
Hence, the redundancy is at most $$4+\log\frac{n}{t-s}+(t-s-1)(\log(\log\frac{n}{t-s}+4)+3),$$
which is $\log n+ (t-s-1)\log\log n+O(1)$.
\end{IEEEproof}

\subsection{Further discussions}

Note that, $(t,s)$-burst correcting codes can correct $(t-s,0)$-burst errors (i.e., $(t-s)$-burst-deletion), then $(t,s)$-burst correcting codes are naturally also $(t-s)$-burst-deletion-correcting codes. Regarding the redundancy, the optimal redundancy of $(t,s)$-burst correcting codes should be lower bounded by the optimal redundancy of $(t-s)$-burst-deletion-correcting codes.

Up till now the best known construction of $(t-s)$-burst-deletion-correcting codes is the one from \cite{Schoeny-Burst-IT17}, where the redundancy is about $\log n+(t-s-1)\log\log n+O(1)$. Comparing with the redundancy of our $(t,s)$-burst correcting codes from Construction \ref{constr2s}, there is only a difference of a constant term. Therefore, as a byproduct, our codes from Construction \ref{constr2s} also performs well against only $(t-s)$-burst-deletions.

Furthermore, note that there is only a $\log\log n$ gap between the redundancy of Construction \ref{constr2s} and the lower bound suggested by the sphere-packing bound.
In the next section, for $t=3$ and $s=1$ we manage to close this gap.

\section{Optimal Codes Correcting a $(3,1)$-Burst}\label{sec:3-1-constr}

A $(2,1)$-burst error, except for the two forms $11\rightarrow 0$, $00\rightarrow 1$, can be seen as a single deletion. The almost optimal $(2,1)$-burst correcting code with redundancy $\log n+3$ from\cite{Schoeny-Burst-IT17}, as shown in Equation (\ref{Equ:2-1-burst}), adds an additional constraint on the basis of the VT code, in order to deal with the errors of the form $11\rightarrow 0$ and $00\rightarrow 1$.
Motivated by this construction and the fact that a large proportion of $(3,1)$-burst errors could be seen as a $2$-burst-deletion, we build our code based on Levenshtein's code for $2$-burst-deletions as shown in Equation (\ref{Equ:Lev2burst}), and add some additional constraints in order to deal with the other kinds of $(3,1)$-burst errors.

\begin{construction}
  For $a\in\mathbb{Z}_{4n}$, $b,c\in\mathbb{Z}_4$, and $d\in\mathbb{Z}_5$, the code $\mathcal{C}_{3,1}$ is defined as follows:
  \begin{equation*}
       \mathcal{C}_{3,1}(n;a,b,c,d)=\Big\{\boldsymbol x: Rsyn(0\boldsymbol x)\equiv a~(\bmod{4n}),
          \sum_{i=1}^{n/2}x_{2i-1}\equiv b~(\bmod4),
         \sum_{i=1}^{n/2}x_{2i}\equiv c~(\bmod4),
          r(\boldsymbol x)\equiv d~(\bmod5)\Big\}.
  \end{equation*}
\end{construction}

In order to show that $\mathcal{C}_{3,1}$ is a $(3,1)$-burst correcting code we need several steps. First we divide all $(3,1)$-burst errors into two types. The first type contains those which can be seen as a $2$-burst, for example, $001\rightarrow 1$, $000\rightarrow 0$. The second type only consists of the following: $\{ 000\rightarrow 1, 010 \rightarrow 1 , 111 \rightarrow 0, 101 \rightarrow 0\}$. To verify which type of error occurs, we need to observe the following two values. Suppose $\boldsymbol u\in\mathbb{F}_2^n$ is a codeword in $\mathcal{C}_{3,1}(n;a,b,c,d)$, and $\boldsymbol u'\in\mathbb{F}_2^{n-2}$ is the sequence obtained from $\boldsymbol u$ by a $(3,1)$-burst.
Let $\Delta_{odd}(\boldsymbol u)=\sum_{i=1}^{n/2}u_{2i-1}-\sum_{i=1}^{n/2-1}u'_{2i-1}\pmod4$ and $\Delta_{even}(\boldsymbol u)=\sum_{i=1}^{n/2}u_{2i}-\sum_{i=1}^{n/2-1}u'_{2i}\pmod4$.

\begin{lemma}\label{lem:oddeven}
By observing $\Delta_{odd}(\boldsymbol u)$ and $\Delta_{even}(\boldsymbol u)$, one can verify whether the $(3,1)$-burst error is a $2$-burst-deletion or is of the form $\{ 000\rightarrow 1, 010 \rightarrow 1 , 111 \rightarrow 0, 101 \rightarrow 0\}$.
\end{lemma}

\begin{IEEEproof}
Suppose the $(3,1)$-burst starts at the $i$th coordinate of $\boldsymbol u=(u_1,u_2,\dots,u_n)$ and results in $(u_1,\dots,u_{i-1},y,u_{i+3},\dots,u_{n})$, where $y$ is the inserted symbol. If $i$ is odd, then $\Delta_{odd}(\boldsymbol u)=u_{i}+u_{i+2}-y$ and $\Delta_{even}(\boldsymbol u)=u_{i+1}$. Similarly, if $i$ is even then $\Delta_{even}(\boldsymbol u)=u_{i}+u_{i+2}-y$ and $\Delta_{odd}(\boldsymbol u)=u_{i+1}$. The correspondence of the error type and $(\Delta_{odd}(\boldsymbol u),\Delta_{even}(\boldsymbol u))$ is as follows:
\begin{itemize}
  \item $000\rightarrow 1$. Then $(\Delta_{odd}(\boldsymbol u),\Delta_{even}(\boldsymbol u))$ is either $(3,0)$ or $(0,3)$.
  \item $010\rightarrow 1$. Then $(\Delta_{odd}(\boldsymbol u),\Delta_{even}(\boldsymbol u))$ is either $(3,1)$ or $(1,3)$.
  \item $111\rightarrow 0$. Then $(\Delta_{odd}(\boldsymbol u),\Delta_{even}(\boldsymbol u))$ is either $(2,1)$ or $(1,2)$.
  \item $101\rightarrow 0$. Then $(\Delta_{odd}(\boldsymbol u),\Delta_{even}(\boldsymbol u))$ is either $(2,0)$ or $(0,2)$.
  \item If the error is a 2-deletion, the two deleted symbols are $00,01,10,11$ and $(\Delta_{odd}(\boldsymbol u),\Delta_{even}(\boldsymbol u))\in \{(0,0),(0,1),(1,0),(1,1)\}$.
\end{itemize}

Thus, the disjointness of the possible values of $(\Delta_{odd}(\boldsymbol u),\Delta_{even}(\boldsymbol u))$ allows us to precisely determine the error type.
\end{IEEEproof}

The first step of our decoding process is to observe $(\Delta_{odd}(\boldsymbol x),\Delta_{even}(\boldsymbol x))$ based on Lemma \ref{lem:oddeven}. Suppose that we have verified that the error is indeed a $2$-burst-deletion.
Then, as proved by Levenshtein\cite{Levenshtein-2Burst-1970}, the constraint $Rsyn(0\boldsymbol x)\equiv a ~ (\bmod{4n})$ in the code $\mathcal{C}_{3,1}$ guarantees that we can successfully decode any $2$-burst-deletion error. Therefore, it suffices to analyze the case when the error is of the form $\{ 000\rightarrow 1, 010 \rightarrow 1 , 111 \rightarrow 0, 101 \rightarrow 0\}$. In the rest of this section we only prove that $\mathcal{C}_{3,1}$ is an error-correcting code against the error type $000\rightarrow 1$ or the error type $010\rightarrow 1$. The error types $111\rightarrow 0$ and $101\rightarrow 0$ can be proved analogously and are thus omitted.

\begin{lemma}\label{lem:000-1}
$\mathcal{C}_{3,1}$ is an error-correcting code against a $000\rightarrow 1$ error.
\end{lemma}

\begin{IEEEproof}
Suppose we have two distinct codewords $\boldsymbol x, \boldsymbol y\in\mathcal{C}_{3,1}(n;a,b,c,d)$ and $\boldsymbol{z}\in \mathbb{F}_2^{n-2}$ can be derived from both $\boldsymbol x$ and $\boldsymbol y$ by a $000\rightarrow 1$ error. Then $\boldsymbol x$ and $\boldsymbol y$ should be of the form
\begin{equation*}
      \begin{split}
        &\boldsymbol x=(x_{1},\ldots,x_{i-1},0,0,0,x_{i+3},\ldots,x_{j+1},1,x_{j+3},\ldots,x_{n}), \\ &\boldsymbol y=(y_{1},\ldots,y_{i-1},1,y_{i+1},\ldots,y_{j-1},0,0,0,y_{j+3},\ldots,y_{n}).
      \end{split}
\end{equation*}
where $x_k=y_k$ when $1\leq k\leq i-1$ and $j+3\leq k\leq n$, $x_{k+2}=y_k$ when $i+1\leq k\leq j-1$.

We turn to the last constraint $r(\boldsymbol x)$ in the definition of $\mathcal{C}_{3,1}$, which is the number of runs. Let $\Delta_r(\boldsymbol x)=r(\boldsymbol x)-r(\boldsymbol z)\pmod 5$ and $\Delta_r(\boldsymbol y)=r(\boldsymbol y)-r(\boldsymbol z)\pmod 5$. Since $\boldsymbol x, \boldsymbol y$ are both codewords from $\mathcal{C}_{3,1}$, we must have $\Delta_r(\boldsymbol x)=\Delta_r(\boldsymbol y)$ and its value is of the following possibilities:
\begin{itemize}
  \item If $i=1$, i.e., the $(3,1)$-burst starts at the very beginning, then depending on whether $x_4=0$ or $x_4=1$, the error pattern is either $0000\rightarrow10$ or $0001\rightarrow11$ and thus $\Delta_r(\boldsymbol x)$ is $4$ or $1$.
  \item If $i=n-2$, i.e., the $(3,1)$-burst starts at the end, then depending on whether $x_{n-3}=0$ or $x_{n-3}=1$, the error pattern is either $0000\rightarrow01$ or $1000\rightarrow11$ and thus $\Delta_r(\boldsymbol x)$ is $4$ or $1$.
  \item Otherwise, consider the four different cases for $x_{i-1}$ and $x_{i+3}$. The error pattern is $00000\rightarrow010, 00001\rightarrow 011, 10000\rightarrow 110, 10001\rightarrow 111$ and the corresponding value of $\Delta_r(\boldsymbol x)$ is $3$, $0$, $0$, and $2$.
\end{itemize}

The rest of the proof falls into five cases. In each case we arrive at a contradiction and thus prove that the assumption of the two distinct codewords $\boldsymbol x, \boldsymbol y\in\mathcal{C}_{3,1}(n;a,b,c,d)$ is not valid.

\begin{itemize}
  \item Case 1: $\Delta_r(\boldsymbol x)=\Delta_r(\boldsymbol y)=1$. Then according to the analysis above, we have
  \begin{equation*}
        \boldsymbol x=(0,0,0,1,x_{5},\ldots,x_{n-2},1,1), ~ \boldsymbol y=(1,1,y_{3},\ldots,y_{n-4},1,0,0,0),
      \end{equation*}
  where $x_{k+2}=y_k$ when $3\leq k\leq n-4$ and the $000\rightarrow 1$ error starts at the first coordinate of $\boldsymbol{x}$ and the $(n-2)$th coordinate of $\boldsymbol{y}$. Now we turn back to the constraint $Rsyn(0\boldsymbol{x})$ for contradiction. The run sequence $R(0\boldsymbol{x})$ is of the form
  $$(0,0,0,0,1)\sim(\text{run sequence corresponding to }x_5,\dots,x_{n-2})\sim(\lambda-1,\lambda-1)$$
  where $\lambda=r(0\boldsymbol{x})=r(\boldsymbol x)$. The run sequence $R(0\boldsymbol{y})$ is of the form
  $$(0,1,1)\sim(\text{run sequence corresponding to } y_3,\dots,y_{n-4})\sim(\lambda'-2,\lambda'-1,\lambda'-1,\lambda'-1)$$
  where $\lambda'=r(0\boldsymbol{y})=r(\boldsymbol y)+1$. Note that for every $3\leq k\leq n-4$, the run index of $x_{k+2}$ in $R(0\boldsymbol{x})$ equals the run index of $y_{k}$ in $R(0\boldsymbol{y})$. Also note that according to the definition of the code we must have $r(\boldsymbol x)=r(\boldsymbol y)$. Thus we have
\begin{equation*}
  Rsyn(0\boldsymbol x)-Rsyn(0\boldsymbol y)= (2\lambda-1)- (4\lambda'-3) = -2r(\boldsymbol x)-2 \neq 0 \pmod{4n},
\end{equation*}
which is a contradiction to the constraint that $Rsyn(0\boldsymbol x)\equiv Rsyn(0\boldsymbol y) \pmod{4n}$.

  \item Case 2: $\Delta_r(\boldsymbol x)=\Delta_r(\boldsymbol y)=4$. Similarly as the previous case, we have
  \begin{equation*}
  \boldsymbol x=(0,0,0,0,x_{5},\ldots,x_{n-2},0,1), ~ \boldsymbol y=(1,0,y_{3},\ldots,y_{n-4},0,0,0,0).
  \end{equation*}
  where $x_{k+2}=y_k$ when $3\leq k\leq n-4$ and the $000\rightarrow 1$ error starts at the first coordinate of $\boldsymbol{x}$ and the $(n-2)$th coordinate of $\boldsymbol{y}$. The run sequence $R(0\boldsymbol{x})$ is of the form
  $$(0,0,0,0,0)\sim(\text{run sequence corresponding to  }x_5,\dots,x_{n-2})\sim(\lambda-2,\lambda-1)$$
  where $\lambda=r(0\boldsymbol{x})=r(\boldsymbol x)$. The run sequence $R(0\boldsymbol{y})$ is of the form
  $$(0,1,2)\sim(\text{run sequence corresponding to  }y_3,\dots,y_{n-4})\sim(\lambda'-1,\lambda'-1,\lambda'-1,\lambda'-1)$$ where $\lambda'=r(0\boldsymbol{y})=r(\boldsymbol y)+1$. Note that for every $3\leq k\leq n-4$, the run index of $x_{k+2}$ in $R(0\boldsymbol{x})$ is the run index of $y_{k}$ in $R(0\boldsymbol{y})$ minus two. Thus we have
\begin{equation*}
  Rsyn(0\boldsymbol x)-Rsyn(0\boldsymbol y)= (2\lambda-3) - 2(n-6) - (4\lambda'-1) = -2n -2r(\boldsymbol x) +6 \pmod{4n}.
\end{equation*}
  From the representation of $\boldsymbol{x}$, we have $2\leq r(\boldsymbol x)\leq n-3$. Thus $-4n+12\leq -2n -2r(\boldsymbol x) +6 \leq -2n+2$ and thus $Rsyn(0\boldsymbol x)-Rsyn(0\boldsymbol y)$ is nonzero modulo $4n$, a contradiction.

  \item Case 3: $\Delta_r(\boldsymbol x)=\Delta_r(\boldsymbol y)=3$ and the error pattern must be $00000\rightarrow 010$. We have
      \begin{equation*}
      \begin{split}
        &\boldsymbol x=(\ldots,\overset{a}0,\overset{a}0,\overset{a}0,\overset{a}0,\overset{a}0,x_{i+4},\ldots,x_{j},\overset{b}0,\overset{b+1}1,\overset{b+2}0,\ldots), \\ &\boldsymbol y=(\ldots,\underset{a}0,\underset{a+1}1,\underset{a+2}0,y_{i+2},\ldots,y_{j-2},\underset{b+2}0,\underset{b+2}0,\underset{b+2}0,\underset{b+2}0,\underset{b+2}0,\ldots),
      \end{split}
      \end{equation*}
  where the $a$'s and $b$'s are the run index of corresponding entries. Moreover, for every $i+2\leq k\leq j-2$, the run index of $x_{k+2}$ in $R(0\boldsymbol{x})$ is the run index of $y_{k}$ in $R(0\boldsymbol{y})$ minus two. In this case,
\begin{equation*}
  Rsyn(0\boldsymbol x)-Rsyn(0\boldsymbol y)= (5a+3b+3) - 2(j-i-3) - (3a+5b+13) = 2(a-b)-2(j-i)-4  \pmod{4n}.
\end{equation*}
  From the representation of $\boldsymbol{x}$ we have $a\leq b\leq a+j-i-2$. Hence, $-4(j-i)\leq Rsyn(0\boldsymbol x)-Rsyn(0\boldsymbol y)\leq -2(j-i)-4$. Moreover, $2\leq j-i\leq n-5$, thus we have $-4(n-5)\leq Rsyn(0\boldsymbol x)-Rsyn(0\boldsymbol y)\leq -8$, which is again a contradiction to the constraint that $Rsyn(0\boldsymbol x)-Rsyn(0\boldsymbol y)\equiv0\pmod{4n}$.

  \item Case 4: $\Delta_r(\boldsymbol x)=\Delta_r(\boldsymbol y)=2$ and the error pattern must be $10001\rightarrow 111$. We have
      \begin{equation*}
      \begin{split}
        &\boldsymbol x=(\ldots,\overset{a}1,\overset{a+1}0,\overset{a+1}0,\overset{a+1}0,\overset{a+2}1,x_{i+4},\ldots,x_{j},\overset{b}1,\overset{b}1,\overset{b}1,\ldots), \\ &\boldsymbol y=(\ldots,\underset{a}1,\underset{a}1,\underset{a}1,y_{i+2},\ldots,y_{j-2},\underset{b-2}1,\underset{b-1}0,\underset{b-1}0,\underset{b-1}0,\underset{b}1,\ldots).
      \end{split}
\end{equation*}
  In this case, for every $i+2\leq k\leq j-2$, the run index of $x_{k+2}$ in $R(0\boldsymbol{x})$ is the run index of $y_{k}$ in $R(0\boldsymbol{y})$ plus two. Then we have
\begin{equation*}
  Rsyn(0\boldsymbol x)-Rsyn(0\boldsymbol y)= (5a+3b+5) + 2(j-i-3) - (3a+5b-5) = 2(a-b)+2(j-i)+4  \pmod{4n}.
\end{equation*}
  From the representation of $\boldsymbol{x}$ we have $a+2\leq b\leq a+j-i$. Hence, $4\leq Rsyn(0\boldsymbol x)-Rsyn(0\boldsymbol y)\leq 2(j-i)$. Moreover, $2\leq j-i\leq n-5$, thus we have $ 4 \leq Rsyn(0\boldsymbol x)-Rsyn(0\boldsymbol y)\leq 2(n-5)$, which is again a contradiction to the constraint that $Rsyn(0\boldsymbol x)-Rsyn(0\boldsymbol y)\equiv0\pmod{4n}$.

  \item Case 5: $\Delta_r(\boldsymbol x)=\Delta_r(\boldsymbol y)=0$. This case is further divided depending on whether each error is $00001\rightarrow 011$ or $10000\rightarrow 110$. We only present one subcase as an example and the others can be proved analogously. Consider the subcase when both the error patterns in $\boldsymbol x$ and $\boldsymbol y$ are $00001\rightarrow 011$. We have
  \begin{equation*}
      \begin{split}
        &\boldsymbol x=(\ldots,\overset{a}0,\overset{a}0,\overset{a}0,\overset{a}0,\overset{a+1}1,x_{i+4},\ldots,x_{j},\overset{b}0,\overset{b+1}1,\overset{b+1}1,\ldots), \\ &\boldsymbol y=(\ldots,\underset{a}0,\underset{a+1}1,\underset{a+1}1,y_{i+2},\ldots,y_{j-2},\underset{b}0,\underset{b}0,\underset{b}0,\underset{b}0,\underset{b+1}1,\ldots),
      \end{split}
  \end{equation*}
  In this subcase, for every $i+2\leq k\leq j-2$, the run index of $x_{k+2}$ in $R(0\boldsymbol{x})$ equals the run index of $y_{k}$ in $R(0\boldsymbol{y})$. Then we have
\begin{equation*}
  Rsyn(0\boldsymbol x)-Rsyn(0\boldsymbol y)= (5a+3b+3) - (3a+5b+3) =2(a-b) \pmod{4n}.
\end{equation*}
In this subcase, $a+2\leq b\leq a+j-i-1$. Therefore, $-2(n-6) \leq-2(j-i-1)\leq  Rsyn(0\boldsymbol x)-Rsyn(0\boldsymbol y)\leq -4$. Moreover, $2 \leq j-i\leq n-5$, $ -2n+12 \leq Rsyn(0\boldsymbol x)-Rsyn(0\boldsymbol y)\leq -4$, which is again a contradiction to the constraint that $Rsyn(0\boldsymbol x)-Rsyn(0\boldsymbol y)\equiv0\pmod{4n}$.
\end{itemize}

To sum up, we have assumed that we have two distinct codewords $\boldsymbol x, \boldsymbol y\in\mathcal{C}_{3,1}(n;a,b,c,d)$ and $\boldsymbol{z}\in \mathbb{F}_2^{n-2}$ can be derived from both $\boldsymbol x$ and $\boldsymbol y$ by a $000\rightarrow 1$ error. However, in all cases we can deduce that  $Rsyn(0\boldsymbol x)-Rsyn(0\boldsymbol y)$ is nonzero modulo $4n$, which contradicts to the definition of the code. Thus, we have proven that $\mathcal{C}_{3,1}$ is an error-correcting code against a $000\rightarrow 1$ error.
\end{IEEEproof}

\medskip

\begin{lemma}\label{lem:010-1}
$\mathcal{C}_{3,1}$ is an error-correcting code against $010\rightarrow 1$ errors.
\end{lemma}

\begin{IEEEproof}
The proof follows the same way as the previous lemma. Suppose that we have two distinct codewords $\boldsymbol x, \boldsymbol y\in\mathcal{C}_{3,1}(n;a,b,c,d)$ and $\boldsymbol{z}\in \mathbb{F}_2^{n-2}$ can be derived from both $\boldsymbol x$ and $\boldsymbol y$ by a $010\rightarrow 1$ error. Then $\boldsymbol x$ and $\boldsymbol y$ should be of the form
\begin{equation*}
      \begin{split}
        &\boldsymbol x=(x_{1},\ldots,x_{i-1},0,1,0,x_{i+3},\ldots,x_{j+1},1,x_{j+3},\ldots,x_{n}), \\ &\boldsymbol y=(y_{1},\ldots,y_{i-1},1,y_{i+1},\ldots,y_{j-1},0,1,0,y_{j+3},\ldots,y_{n}).
      \end{split}
\end{equation*}
where $x_k=y_k$ when $1\leq k\leq i-1$ and $j+3\leq k\leq n$, $x_{k+2}=y_k$ when $i+1\leq k\leq j-1$.

First we need to observe the two values $\Delta_r(\boldsymbol x),\Delta_r(\boldsymbol y)$.

\begin{itemize}
  \item If $i=1$, i.e., the $(3,1)$-burst starts at the very beginning, then depending on whether $x_4=0$ or $x_4=1$, the error pattern is either $0100\rightarrow10$ or $0101\rightarrow11$ and thus $\Delta_r(\boldsymbol x)$ is $1$ or $3$.
  \item If $i=n-2$, i.e., the $(3,1)$-burst starts at the end, then depending on whether $x_{n-3}=0$ or $x_{n-3}=1$, the error pattern is either $0010\rightarrow01$ or $1010\rightarrow11$ and thus $\Delta_r(\boldsymbol x)$ is $1$ or $3$.
  \item Otherwise, consider the four different cases for $x_{i-1}$ and $x_{i+3}$. The error pattern is $00100\rightarrow010, 00101\rightarrow 011, 10100\rightarrow 110, 10101\rightarrow 111$ and the corresponding value of $\Delta_r(\boldsymbol x)$ is $0$, $2$, $2$, and $4$.
\end{itemize}

Based on the observations of $\Delta_r(\boldsymbol x),\Delta_r(\boldsymbol y)$ we then break into several cases and in each case we will derive a contradiction by analyzing $Rsyn(0\boldsymbol x)-Rsyn(0\boldsymbol y)$. Since the whole framework is similar as the previous lemma, we omit some details and only present the calculations for $Rsyn(0\boldsymbol x)-Rsyn(0\boldsymbol y)$.

\begin{itemize}
  \item Case 1: $\Delta_r(\boldsymbol x)=\Delta_r(\boldsymbol y)=1$.
  \begin{equation*}
        \boldsymbol x=(0,1,0,0,x_{5},\ldots,x_{n-2},0,1), ~ \boldsymbol y=(1,0,y_{3},\ldots,y_{n-4},0,0,1,0).
  \end{equation*}
  Then $Rsyn(0\boldsymbol x)-Rsyn(0\boldsymbol y)=-2r(\boldsymbol x)+4\neq0$, since $r(\boldsymbol x)\geq 4$.

  \item Case 2: $\Delta_r(\boldsymbol x)=\Delta_r(\boldsymbol y)=3$.
  \begin{equation*}
        \boldsymbol x=(0,1,0,1,x_{5},\ldots,x_{n-2},1,1), ~ \boldsymbol y=(1,1,y_{3},\ldots,y_{n-4},1,0,1,0).
  \end{equation*}
  This case is a little bit special and deserves to be analyzed in detail. We can compute that $Rsyn(0\boldsymbol x)-Rsyn(0\boldsymbol y)=2n-2r(\boldsymbol x)-4$. Suppose it is zero modulo $4n$, then $n=r(\boldsymbol x)+2$. Since the first symbol of $\boldsymbol x$ is $0$ and the last symbol of  $\boldsymbol x$ is $1$, then the number of runs $r(\boldsymbol x)$ must be even and thereby $n$ must be even. As $\sum_{k=1}^{n/2}x_{2k-1}=1+\sum_{k=3}^{n/2-1}x_{2k-1}$, $\sum_{k=1}^{n/2}y_{2k-1}=3+\sum_{k=2}^{n/2-2}y_{2k-1}$ and $\sum_{k=3}^{n/2-1}x_{2k-1}=\sum_{k=2}^{n/2-2}y_{2k-1}$, we can get $\sum_{k=1}^{n/2}x_{2k-1}-\sum_{k=1}^{n/2}y_{2k-1}\equiv2\pmod4$ which is a contradiction to  $\sum_{k=1}^{n/2}x_{2k-1}-\sum_{k=1}^{n/2}y_{2k-1}\equiv0\pmod4$. Therefore, $Rsyn(0\boldsymbol x)-Rsyn(0\boldsymbol y)=2n-2r(\boldsymbol x)-4\neq0 \pmod{4n}$.

  \item Case 3: $\Delta_r(\boldsymbol x)=\Delta_r(\boldsymbol y)=0$ and the error pattern must be $00100\rightarrow010$. Hence,
\begin{equation*}
      \begin{split}
        &\boldsymbol x=(\ldots,\overset{a}0,\overset{a}0,\overset{a+1}1,\overset{a+2}0,\overset{a+2}0,x_{i+4},\ldots,x_{j},\overset{b}0,\overset{b+1}1,\overset{b+2}0,\ldots), \\ &\boldsymbol y=(\ldots,\underset{a}0,\underset{a+1}1,\underset{a+2}0,y_{i+2},\ldots,y_{j-2},\underset{b}0,\underset{b}0,\underset{b+1}1,\underset{b+2}0,\underset{b+2}0,\ldots),
      \end{split}
\end{equation*}
Then $Rsyn(0\boldsymbol x)-Rsyn(0\boldsymbol y)=2(a-b)\neq 0 \pmod{4n}.$

  \item Case 4: $\Delta_r(\boldsymbol x)=\Delta_r(\boldsymbol y)=4$ and the error pattern must be $10101\rightarrow111$. Hence,
\begin{equation*}
      \begin{split}
        &\boldsymbol x=(\ldots,\overset{a}1,\overset{a+1}0,\overset{a+2}1,\overset{a+3}0,\overset{a+4}1,x_{i+4},\ldots,x_{j},\overset{b}1,\overset{b}1,\overset{b}1,\ldots), \\ &\boldsymbol y=(\ldots,\underset{a}1,\underset{a}1,\underset{a}1,y_{i+2},\ldots,y_{j-2},\underset{b-4}1,\underset{b-3}0,\underset{b-2}1,\underset{b-1}0,\underset{b}1,\ldots),
      \end{split}
\end{equation*}
Then $Rsyn(0\boldsymbol x)-Rsyn(0\boldsymbol y)=2(a-b)+4(j-i)+8.$ Since $a+4\leq b\leq a+j-i+2$, we have $2(j-i)+4 \leq Rsyn(0\boldsymbol x)-Rsyn(0\boldsymbol y)\leq4(j-i)$, so $Rsyn(0\boldsymbol x)-Rsyn(0\boldsymbol y)\neq0 \pmod{4n}$.

  \item Case 5: $\Delta_r(\boldsymbol x)=\Delta_r(\boldsymbol y)=2$.
  his case is further divided depending on whether each error is $00101\rightarrow 011$ or $10100\rightarrow 110$. We only present one subcase as an example and the others can be proved analogously. Consider the subcase when the error pattern in $\boldsymbol x$ is $00101\rightarrow011$
      and the error pattern in $\boldsymbol y$ is $10100\rightarrow110$. We have
\begin{equation*}
      \begin{split}
        &\boldsymbol x=(\ldots,\overset{a}0,\overset{a}0,\overset{a+1}1,\overset{a+2}0,\overset{a+3}1,x_{i+4},\ldots,x_{j},\overset{b}1,\overset{b}1,\overset{b+1}0,\ldots), \\ &\boldsymbol y=(\ldots,\underset{a}0,\underset{a+1}1,\underset{a+1}1,y_{i+2},\ldots,y_{j-2},\underset{b-2}1,\underset{b-1}0,\underset{b}1,\underset{b+1}0,\underset{b+1}0,\ldots),
      \end{split}
\end{equation*}
Then $Rsyn(0\boldsymbol x)-Rsyn(0\boldsymbol y)=2(a-b)+2(j-i)$. Suppose it is zero, then $r_j-r_i=j-i$. Since the symbols with run index $a$ are $0$s and symbols with run index $b$ are $1$s, $b-a$ must be odd and thus $j-i$ is odd. The number of symbols from $x_{i+4}$ to $x_j$ is $j-i-3$ and thus is even. WLOG let $i$ be odd, then similarly as Case 2 we can analyze the sum of the odd coordinates and derive $$\sum_{k=1}^{n/2}x_{2k-1}- \sum_{k=1}^{n/2}y_{2k-1}\equiv1-3\equiv2\pmod4,$$ which is a contradiction to $\sum_{k=1}^{n/2}x_k-\sum_{k=1}^{n/2}y_k\equiv0\pmod4$. As a consequence, $Rsyn(0\boldsymbol x)-Rsyn(0\boldsymbol y)\neq0 \pmod{4n}$.
\end{itemize}

To sum up, we have assumed that we have two distinct codewords $\boldsymbol x, \boldsymbol y\in\mathcal{C}_{3,1}(n;a,b,c,d)$ and $\boldsymbol{z}\in \mathbb{F}_2^{n-2}$ can be derived from both $\boldsymbol x$ and $\boldsymbol y$ by a $010\rightarrow 1$ error. However, in all cases we can deduce that  $Rsyn(0\boldsymbol x)-Rsyn(0\boldsymbol y)$ is nonzero modulo $4n$, which contradicts to the definition of the code. Thus, we have proven that $\mathcal{C}_{3,1}$ is an error-correcting code against a $010\rightarrow 1$ error.
\end{IEEEproof}

We close this section by summarizing our construction of the code $\mathcal{C}_{3,1}(n;a,b,c,d)$:
\begin{itemize}
  \item The decoding process starts with the observations of $\Delta_{odd}(\boldsymbol u)$ and $\Delta_{even}(\boldsymbol u)$.
  \item If we determine that the error can be seen as a $2$-burst-deletion, then according to Levenshtein\cite{Levenshtein-2Burst-1970} our code is capable of correcting a $2$-burst-deletion error.
  \item Otherwise, we can determine the error pattern, which is one out of $\{000\rightarrow 1, 010 \rightarrow 1 , 111 \rightarrow 0, 101 \rightarrow 0\}$.
  \item Whatever the error pattern is, our code can correct this type of error (two models are proved via Lemmas \ref{lem:000-1} and \ref{lem:010-1} and the other two follow a similar idea).
\end{itemize}

Finally, by the pigeonhole principle, we may find suitable parameters $a\in\mathbb{Z}_{4n}$, $b,c\in\mathbb{Z}_4$, and $d\in\mathbb{Z}_5$ and find a code with size at least $\frac{2^n}{4n\cdot 80}$ and thus its redundancy is at most $\log(320n)<\log n+9$. Note that we have proved that the lower bound of the redundancy of $(3,1)$-burst correcting codes is $\log n+2$, so our construction is optimal up to a constant. In sum, in this section we have proved the following.

\begin{theorem}
There exist choices of $a\in\mathbb{Z}_{4n}$, $b,c\in\mathbb{Z}_4$, and $d\in\mathbb{Z}_5$, such that the code $\mathcal{C}_{3,1}(n;a,b,c,d)$ is a $(3,1)$-burst correcting code with redundancy at most $\log n+9$.
\end{theorem}

\section{Conclusion and Future Work}\label{sec:concl}
In this paper we study $(t,s)$-burst correcting codes. First we prove the equivalence between $(t,s)$-burst correcting codes and $(s,t)$-burst correcting codes. Then we present a sphere-packing type upper bound of $(t,s)$-burst correcting codes, leading to a lower bound of its optimal redundancy. We present a construction of $(t,s)$-burst correcting codes for $t\geq2s$, with redundancy $\log n+(t-s-1)\log\log n+O(1)$. Comparing our general construction and the lower bound of redundancy, there is only a $\log\log n$ gap. We manage to close this gap for $t=3,s=1$ by giving a construction of $(3,1)$-burst correcting codes with redundancy at most $\log n+9$.

Here is a brief remark on our general construction for $t\geq 2s$. If we want to generalize the construction for the range $s\leq t <2s$, then by viewing the codeword as an array of size $(t-s)\times\frac{n}{t-s}$, each row will suffer from more than a $(2,1)$-burst. To be more specific, if $\frac{p-2}{p-1}< \frac{s}{t}\leq \frac{p-1}{p}$, then in the array representation a row may suffer from a $(p,p-1)$-burst. Therefore, good constructions for codes against a $(p,p-1)$-burst for $p\geq 3$ will be useful tools in analyzing the general problem of $(t,s)$-burst correcting codes.

Another interesting question is to construct a code capable of correcting any $(t,s)$-burst for $t\leq T$ and $s\leq S$, with $T$ and $S$ given. Despite a trivial solution to intersect all $(t,s)$-burst correcting codes for $t\leq T$ and $s\leq S$, a nontrivial solution is considered for future research.

\end{document}